\documentclass[aps,twocolumn,preprintnumbers,showpacs,showkeys,nofootinbib%
]{revtex4}
\usepackage{epsfig}
\usepackage{amssymb,amsmath,amsfonts,amsthm,graphicx,psfrag}
\graphicspath{{./Figures/}}                 % location for color  fig.
\newcommand{\noi}{\noindent}
\newcommand{\beq}{\begin{equation}}
\newcommand{\eeq}{\end{equation}}
\newcommand{\bea}{\begin{eqnarray}}
\newcommand{\eea}{\end{eqnarray}}

\newcommand{\Tab}[1]{Table~\ref{#1}}

\newcommand{\tr}{\operatorname{Tr}}

\begin{document}
\preprint{ITEP-LAT/2011-3}

\title{ Thermal Abelian monopoles as selfdual dyons.
}

\author{V.~G.~Bornyakov}
%\email[]{bornvit@gmail.com}
\affiliation{High Energy Physics Institute, 142280 Protvino, Russia \\
and Institute of Theoretical and Experimental Physics, 117259 Moscow, Russia}

\author{V.~V.~Braguta}
%\email[]{}
\affiliation{High Energy Physics Institute, 142280 Protvino, Russia}

\begin{abstract}
The properties of the thermal Abelian monopoles are studied in the deconfinement phase of
the $SU(2)$ gluodynamics. To remove effects of Gribov copies the simulated annealing
algorithm is applied to fix the maximally Abelian gauge.  To study monopole profile we complete
the first computations of
excess of the nonabelian action density as a function of the distance from the center of
the thermal Abelian monopole. We have found that starting from the distances $\approx 2$ lattice
spacings the chromoelectric and chromomagnetic  action densities created by monopole are equal to each
other, from what we draw a conclusion that monopole is a dyon. Furthermore, we find that the
chromoelectric and chromomagnetic fields decrease exponentially with increasing distance.
These findings were confirmed for different temperatures in the range $T/T_c \in (1.5, 4.8)$.
\end{abstract}



\maketitle

One of the hypotheses which have been put forward in the recent past is that the quark-gluon plasma (QGP)
properties might be
dominated by a magnetic component~\cite{Liao:2006ry,Chernodub:2006gu,Shuryak:2008eq}.
The monopoles or center vortices might be responsible for unexpected properties of
the hadron matter at $T>T_c$: on one hand it is well known from lattice results that the equation of state is close to
that of an ideal gas, on the other hand the very low viscosity to entropy ratio tells that it
is an ideal liquid.

In Ref.~\cite{Chernodub:2006gu} such magnetic component has been related
to thermal Abelian monopoles evaporating from the magnetic condensate which is
believed to induce color confinement at low temperatures.
Moreover it has been proposed to detect such thermal monopoles in
finite temperature lattice QCD simulations, by identifying them
with monopole currents having a non-trivial wrapping in the Euclidean
temporal direction~\cite{Chernodub:2006gu,Bornyakov:1991se,Ejiri:1995gd}.

The way one can study the monopoles properties on
the lattice is via an Abelian projection after fixing the maximally
Abelian gauge (MAG) \cite{'tHooft:1981ht,'tHooft:1982ns,Kronfeld:1987vd}. This gauge as well as the properties of the monopole
clusters has been investigated in a numerous papers both at zero and nonzero
temperature (see for extensive list of references, e.g. \cite{Ripka:2003vv}).
The evidence was found that the nonperturbative properties of the gluodynamics such as confinement,
deconfining transition, chiral symmetry breaking, etc.  are closely related to the Abelian monopoles defined in MAG.
This was called a monopole dominance.

First numerical investigations
of the wrapping  monopole trajectories were performed long ago in Refs.~\cite{Bornyakov:1991se}
and~\cite{Ejiri:1995gd}.  A more systematic study of the thermal monopoles in $SU(2)$ Yang-Mills theory at high temperature
has been performed in Refs.~\cite{D'Alessandro:2007su,Chernodub:2009hc,D'Alessandro:2010xg}. In particular, it was found
in ~\cite{D'Alessandro:2007su} that the density of monopoles
is independent of the lattice spacing, as it should be for a physical quantity.

 In paper \cite{Chernodub:2009hc} very interesting properties of
the thermal Abelian monopoles were found. The authors measured the excess of chromoelectric and chromomagnetic action
density on the surface of the lattice (hyper-)cubes with monopoles inside.
The dependence
of the excess on the distance from the monopole center was determined
through the variation of the lattice spacing $a$ (similar investigation at $T=0$ was made in~\cite{Bornyakov:2001hs}).
As a result it was found that with good accuracy the   chromomagnetic and chromoelectric action
densities created by
monopole have the following  behavior: $H^2(r), E^2(r) = a_{H,E}/r^4$.
The coefficients $a_H$ and $a_E$ turned out to be equal to each other with a very good
accuracy, from what the authors concluded that monopole is a dyon.
It is worth to note that there were other works in the past where dyonic properties of the monopoles were
observed \cite{Chernodub:1995tt,Bornyakov:1996wp,Ilgenfritz:2006ju}.

The drawback of the study undertaken in \cite{Chernodub:2009hc} is that all results were obtained at the ultraviolet cutoff scale 
and were thus subjected to both lattice discretization errors and ultraviolet divergences.
In view of the importance of the findings of \cite{Chernodub:2009hc} in this paper we are
going to study the chromoelectric and chromomagnetic fields created by monopole and
to check whether the observed behavior is correct or is just a lattice artifact, i.e. artifact of the ultraviolet cut off.
To accomplish this check we will measure
the chromoelectric and chromomagnetic fields at various distances from the monopole center.

In this paper we study the SU(2) lattice gauge theory with the standard Wilson action

\beq
S  = \beta \sum_x\sum_{\mu >\nu}
\left[ 1 -\frac{1}{2}\tr \Bigl(U_{x\mu}U_{x+\mu;\nu}
U_{x+\nu;\mu}^{\dagger}U_{x\nu}^{\dagger} \Bigr)\right], \nonumber
\label{eq:action}
\eeq

\noi where $\beta = 4/g_0^2$ and $g_0$ is a bare coupling constant. The
link variables $U_{x\mu} \in SU(2)$ transform  under gauge
transformations $g_x$ as follows:

\beq
U_{x\mu} \stackrel{g}{\mapsto} U_{x\mu}^{g}
= g_x^{\dagger} U_{x\mu} g_{x+\mu} \; ;
\qquad g_x \in SU(2) \,.
\label{eq:gaugetrafo}
\eeq

\noi Our calculations were performed on the asymmetric lattices with
lattice volume $V=L_t L_s^3$, where $L_{t,s}$ is the number of sites in
the time (space) direction. The temperature $T$ is given by

\beq
T = \frac{1}{aL_t}~,
\eeq

\noi where $a$ is the lattice spacing.

The MAG is fixed by finding an extremum of the gauge functional

\beq
F_U(g) = ~\frac{1}{4V}\sum_{x\mu}~\frac{1}{2}~\tr~\biggl( U^{g}_{x\mu}\sigma_3 U^{g\dagger}_{x\mu}\sigma_3 \biggr) \;,
\label{eq:gaugefunctional}
\eeq

\noi with respect to gauge transformations $g_x$. We apply the simulated annealing (SA) algorithm which proved to be very efficient for this
gauge \cite{Bali:1996dm} as well as for other gauges such as center gauges \cite{Bornyakov:2000ig} and Landau gauge \cite{Bogolubsky:2007bw}.
To further decrease the Gribov copy effects we generated 10 Gribov copies starting every time gauge fixing procedure
from a randomly selected gauge copy of the original Monte Carlo configuration.

In \Tab{tab:statistics} we provide the information about the gauge field ensembles used in our study.

%%%%%%%%%%%%%%%%%%%%%%%%%%%%%%%%%%%%%%%%%%%%%%%%%%%%%%%%%%%%%%
%%%%%%%%%%%%%%%%%%%%%%  Table 1  %%%%%%%%%%%%%%%%%%%%%%%%%%%%%
%%%%%%%%%%%%%%%%%%%%%%%%%%%%%%%%%%%%%%%%%%%%%%%%%%%%%%%%%%%%%%

\begin{table}[ht]
\begin{center}
\begin{tabular}{|c|c|c|c|c|c|} \hline
 $\beta$ & $a$[fm] & $L_t$ & $~L_s~$ & $T/T_c$ &
$N_{meas}$ \\ \hline\hline
  2.43   & 0.108 & 4  & 32  & 1.5     & 1000    \\
  2.5115  & 0.081 & 4  & 28  & 2.0     & 400      \\
%  2.55    &            &  4 & 28   & 2.3    & 100   \\
 % 2.60    &           &  4  & 32   & 2.7     &  100 \\
  2.635   & 0.054 & 4  & 36  & 3.0     & 500      \\
  2.80     & 0.034 & 4  & 48  & 4.8     & 400      \\ \hline
\end{tabular}
% }
\end{center}
\caption{Values of $\beta$, lattice sizes, temperatures, number of
measurements and number of gauge copies used throughout this paper.
To fix the scale we take $\sqrt{\sigma}=440$ MeV.
%1 Gev${}^{-1}\simeq 0.1973$ fm.
}
\label{tab:statistics}
\medskip \noindent
\end{table}

%%%%%%%%%%%%%%%%%%%%%%%%%%%%%%%%%%%%%%%%%%%%%%%%%%%%%%%%%%%%%%
%%%%%%%%%%%%%%%%%%%%%%  End Table 1  %%%%%%%%%%%%%%%%%%%%%%%%%
%%%%%%%%%%%%%%%%%%%%%%%%%%%%%%%%%%%%%%%%%%%%%%%%%%%%%%%%%%%%%%

The chromomagnetic action density at a site $x$ is defined as
\beq
 S_M (x)  = \frac{1}{12} \sum_{P_{s} \ni x} \biggl ( 1- \frac 1 2 \tr U_{P_{s}} \biggr ) .
\label{actm}
\eeq
The sum is taken over all spatial plaquettes $P_{s}$ which contain the lattice site $x$.
In the continuum limit this expression is proportional to
$\sim \tr (G_{23}^2+G_{13}^2+G_{12}^2)= \tr (H_1^2+H_2^2+H_3^2)$. So, this
expression can be taken as a measure of the chromomagnetic action $\tr({\bf H}^2)$ at the
site $x$.

Analogously, for the chromoelectric action density at a site $x$ we take
\beq
S_E (x)  = \frac{1}{12} \sum_{P_{t} \ni x} \biggl ( 1- \frac 1 2 \tr U_{P_{t}} \biggr )\,.
\label{acte}
\eeq
Here the sum is taken over all time-like plaquettes $P_{t}$ which contain the site $x$.
In the continuum limit this expression is proportional to
$\sim \tr (G_{01}^2+G_{02}^2+G_{03}^2)= \tr (E_1^2+E_2^2+E_3^2)$ and thus it can be taken as a
measure of the chromoelectric action $\tr({\bf E}^2)$.
Note that our definitions for  $ S_{M,E} (x) $ differ from those used in
Ref.~\cite{Chernodub:2009hc}. Although  definitions of \cite{Chernodub:2009hc}  were natural
for the surface of a cube with monopole our definitions are more suitable for measurements at some distance from such cube.

Since we are studying the fields created by a monopole we should
subtract the vacuum fluctuations of the chromomagnetic and chromoelectric actions
from the equations (\ref{actm}), (\ref{acte}). We define the excess of the action density as
\beq
 \langle\delta S_{M,E} (d)  \rangle= \langle \overline{S_{M,E} (x)} \rangle - \langle S_{M,E} \rangle_{}\,,
\label{actmes}
\eeq
where $\langle ... \rangle$ means ensemble average, $d$ is the distance from a monopole and
bar means averaging over all wrapped monopoles
and all sites $x$ at the distance $d$ from monopole  centers.

The monopole currents and their wrapping numbers are defined in a standard way (see e.g. ~\cite{D'Alessandro:2007su}).
Moving along  wrapped monopole clusters on a dual lattice
we detect all 3-dimensional cubes in all time slices on the original lattice which contain monopoles
 corresponding to $j_4$ component of the magnetic current.
Having detected all such 3-dimensional cubes  we do the measurements of the
chromomagnetic and chromoelectric action densities $  \langle \delta S_{M,E} (d) \rangle $ at various distances
from a given monopole and then we take the average over all thermal monopoles found on the lattice.

Now let us consider a three dimensional cube with the monopole belonging to a wrapped cluster. Below it will be assumed that
the monopole is located in the center of this cube and
we take the center as a coordinate origin. Actually, one cannot assert that
the monopole is exactly located at the center of the cube. However, since we take an average
over all monopoles, this approximation can be considered as a good one. We have measured
the action densities $\langle \delta S_{M,E}(d) \rangle$ at various distances $d$ from the centers of the cubes with monopole.
The distance was defined as a length of the vector $\vec{d} = \{ n_1\pm 1/2, n_2\pm 1/2, n_3\pm 1/2  \}$
from the coordinate origin to lattice site $x$ under consideration.
We present results of  measurements for
$\vec{d} = \frac{1}{2}\{m, m, m\}, m=1,3$ and $\vec{d} = \frac{1}{2}\{1,1,3\}$,  $\frac{1}{2}\{1,3,3\}$,  $\frac{1}{2}\{1,1,5\}$,  $\frac{1}{2}\{1,3,5\}$.
For  $T/T_c=1.5$  additionally results for $\vec{d} = \frac{1}{2}\{1,5,5\}, \frac{1}{2}\{1,1,7\}$ are presented.
For longer distances the statistical errors were too large.

\begin{figure}[tb]
%\vspace*{1.5cm}
\centering
\includegraphics[width=6.0cm,angle=270]{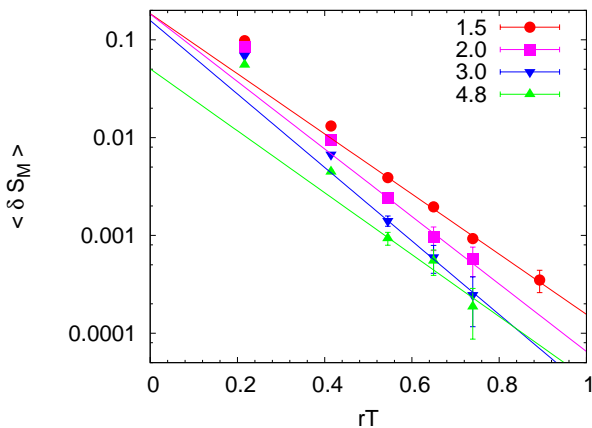}
\caption{$\langle \delta  S_{M} \rangle $ defined in eq.(\ref{actmes}) as function of the dimensionless distance $r T$
from the monopole center at the temperatures $T/T_c=1.5, 2.0, 3.0, 4.8$.
}
\label{fig:mall}
\end{figure}
\begin{figure}[tb]
%\vspace*{1.5cm}
\centering
\includegraphics[width=6.0cm,angle=270]{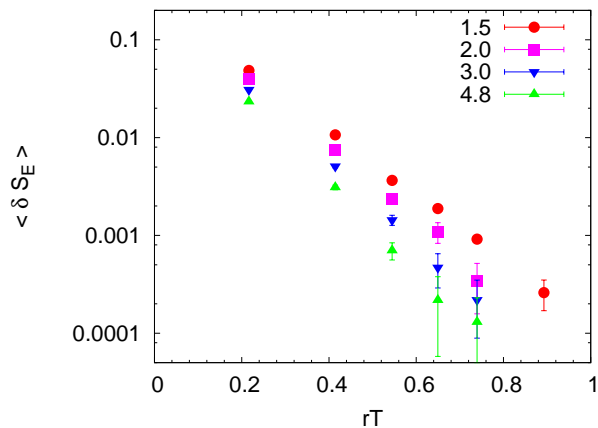}
\caption{Same as in Figure \ref{fig:mall} for $\langle \delta  S_{E} \rangle $.
}
\label{fig:eall}
\end{figure}

In Figures~\ref{fig:mall} and \ref{fig:eall}  results are shown for the  $\langle \delta S_{M} \rangle$  and  $\langle \delta S_{E} \rangle$
as functions of the dimensionless distance $r T = d/4$.
From these Figures
 one clearly sees that at least at large distances both $\langle \delta S_{M} \rangle$  and $\langle \delta S_{E} \rangle$
decrease with distance in agreement with exponential fall-off  $\sim \exp  { (- 2 M_{m,e} r)}$. The dependence $1/r^4$
found in \cite{Chernodub:2009hc} is ruled out.
We do not have enough data points  to determine the pre-exponential function  by fitting. Respectively, it is rather difficult to find
the parameters $M_{m,e}$ with a good accuracy. In this paper we just make rough estimation of these 
parameters
fitting the last 3 data points for $\langle \delta S_{M} \rangle$  to the exponential fall-off  
with constant  prefactor. 
We get the following results: $M_m/T=3.5(2), 4.0(4), 4.3(2) , 3.7(8)$ 
for the temperatures
$T/T_c=1.5, 2.0, 3.0, 4.8$, respectively.

Looking at Figures \ref{fig:mall}, \ref{fig:eall} one can see that the data lie on the smooth curves.
This means that the data obey rotational invariance, since the data at different distances
were measured in different directions. Moreover, vectors  $\vec{d} = \frac{1}{2}\{3, 3, 3\}$ and
$\vec{d} = \frac{1}{2}\{1,1,5\}$ have equal length and one can check the rotational invariance directly.
Indeed, we find for these two vectors consistent results with deviations within $2 \sigma$  interval. In all Figures we
show averaged data for these vectors $\vec{d}$.

\begin{figure}[tb]
%\vspace*{1.5cm}
\centering
\includegraphics[width=6.0cm,angle=270]{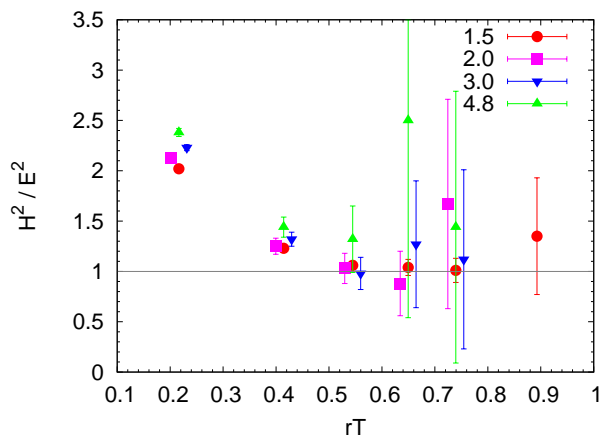}
\caption{The ratio $  \langle  \delta  S_{M} \rangle  / \langle \delta  S_{E} \rangle = H^2/E^2 $ as function of the distance $r T$
from the monopole center at temperatures  $T/T_c=1.5, 2.0, 3.0, 4.8$  The data sets at $T/T_c=2.0$ and  3.0 are shifted 
along horizontal axis to improve readability of the figure.
}
\label{fig:rall}
\end{figure}

At large enough distances ( beginning from the distance $ d=2.18$ ) the monopole 
chromomagnetic and chromoelectric action density
seem to be equal to each other.
To demonstrate this important property we plot the ratio  $ \langle \delta  S_{M} \rangle/ \langle \delta  S_{E} \rangle = H^2/E^2$ 
 in Figure~\ref{fig:rall}.
From this Figure we see that within the error bars at  distances
 $ rT \gtrsim 0.5$   the ratio $\langle \delta  S_{M} \rangle/ \langle \delta  S_{E} \rangle$
is compatible with 1. From this observation one can draw
a conclusion: at least at large enough distances $H^2(r)=E^2(r)$. 
This implies that { \it monopoles carry both chromoelectric and chromomagnetic charges} and 
{\it they are equal}. So {\it monopoles are selfdual dyons. } These statements are the main results of this paper.

To understand the reason of lack of selfduality at small distances let us look at Figure \ref{fig:rall}.
For all temperatures we observe similar behavior. At the distance $d=0.87$   in lattice units
the ratio $\langle \delta  S_{M} \rangle / \langle \delta  S_{E} \rangle \sim 2$ for all temperatures.
At the distance $d=1.66$ the ratio $\langle \delta  S_{M} \rangle / \langle \delta  S_{E} \rangle
\sim 1.3$.
At larger distances the ratio is compatible with unity. We believe that deviation of the ratio from one at small
distances 
can be explained by the discretization  effects. Notice that our definition of the chromomagnetic and chromoelectric
action densities at site $x$ is nonlocal involving all plaquettes in respective planes which own site $x$.
This nonlocality is different for two action densities: it is purely spatial for the  chromomagnetic action density
and is both spatial and temporal for the chromoelectric one. Thus
at distances of order of one lattice spacing
we, evidently, measure the fields taken at different points and different distances.
These our arguments should be checked by computations with smaller lattice spacing, i.e. with $L_t>4$. This will be done
in a forthcoming paper. In that paper we will also present our data on the density and interactions of the thermal Abelian 
monopoles.

It is clear that results and conclusions of Ref.~\cite{Chernodub:2009hc} where $ \langle  \delta  S_{M} \rangle $  and
$  \langle  \delta  S_{E}  \rangle  $ were measured at the nearest possible distance to the
monopole center are subjected to same discretization effects
as discussed above for our data at small distances. While the distance dependence $1/r^4$ found in 
Ref.~\cite{Chernodub:2009hc}  is an ultraviolet divergence effect.

Thus we established that the Abelian thermal  monopoles carry both chromoelectric and chromomagnetic charges and 
they are equal. So  monopoles are selfdual dyons. Furthermore respective action densities are screened.  
We believe that these results are important  for 
understanding QCD in the quark-gluon plasma phase since many recent theoretical models
of this phase include monopoles as an important ingredient. These are model of Ref.~\cite{Liao:2006ry,Shuryak:2008eq} based
on competition between magnetic and electric quasiparticles, dyon  \cite{Diakonov:2007nv} and caloron \cite{Gerhold:2006sk} 
models. 

We would like to express our gratitude to M.I. Polikarpov and V.I. Zakharov for very useful and illuminating discussions.
This investigation has been partly supported by the Federal Special-Purpose Programme 'Cadres' of the Russian Ministry
of Science and Education and partly by the grant for scientific schools NSh-6260.2010.2. VVB is supported by grant
RFBR 10-02-00061. VGB is supported by grants RFBR 09-02-00338-a and RFBR 11-02-01227-a.

\end{document}